\documentstyle[11pt,newpasp,twoside,epsf]{article}
\begin{document}
\title{Red Giants in ISOGAL:\\ Tracers of the Evolution of the Galaxy}
\author{Jacco Th. van Loon}
\affil{Institute of Astronomy, Madingley Road, Cambridge CB3 0HA, UK}
\author{for The ISOGAL Collaboration}
\begin{abstract}

The DENIS/ISOGAL near/mid-IR survey of the Milky Way for the first time probes
stellar populations in the innermost obscured regions of our galaxy. Ages,
metallicities and extinction-corrected luminosities are derived for these
stars individually by isochrone fitting to the preliminary IR photometry. A
few hundred AGB stars with thick circumstellar dust envelopes are identified.
An old metal-rich population dominates in the inner galactic Bulge, but there
are indications for the presence of a younger population. The inner Bulge has
a tri-axial shape, as traced by depth effects on the observed luminosity
distributions.

\end{abstract}

\section{Introduction}

Attempts to understand the evolution of galaxies by studies of their stellar
populations as a function of redshift are limited by sensitivity and angular
resolution. Our nearest galaxy, the Milky Way, has the potential to provide
many important clues on the evolution of galaxies: ages and metallicities may
be measured for individual stars, and the spatial and kinematic distributions
of the different stellar populations may be observed. Astrophysical problems
such as the evolution and mass loss of Asymptotic Giant Branch (AGB) stars,
which are relevant for the (chemical) evolution of galaxies, may be addressed.
For the innermost parts of the Milky Way, however, where most stellar light
and mass are and where most activity is happening, this has not yet been the
case. Due to the location of the Sun in the galactic Plane, our view of the
central few hundred pc of the galactic Bulge and the inner few kpc of the
galactic Disk is obscured by tens of magnitudes extinction at visual
wavelengths.

\begin{figure}
\plotfiddle{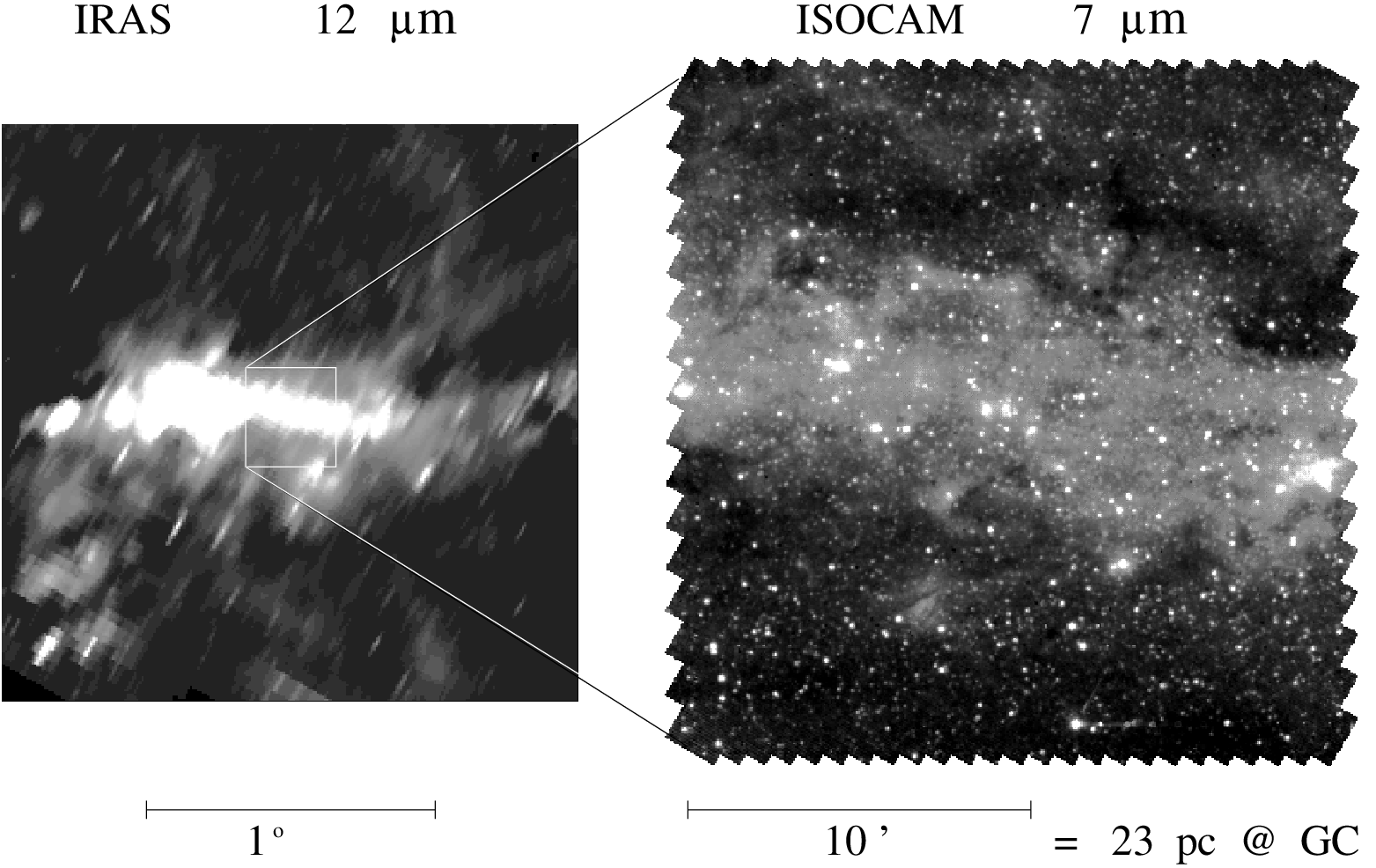}{70mm}{0}{51}{51}{-171}{-10}
\caption{IRAS 12 $\mu$m and ISOGAL 7 $\mu$m images in the direction towards
$(l_{\rm II},b_{\rm II})=(-0.27^{\deg},-0.03^{\deg})$. Clearly, the ISOGAL
data are superior by a few orders of magnitude both in spatial resolution and
sensitivity. Virtually all IR point sources are red giants in the inner Bulge.
Optical images do not show any of these, and are instead severely crowded with
foreground stars.}
\end{figure}

Recent near- and mid-IR surveys of the Milky Way, both from the ground (DENIS,
2MASS, TMGS) and from space (ISOGAL, MSX), for the first time probe large
numbers of stars located deep within the obscured regions of our Galaxy (Fig.\
1). The preliminary catalogue of IR point sources from the DENIS/ISOGAL
0.8--15 $\mu$m survey (Omont et al.\ 1999; see also the contributions of Omont
et al.\ and Blommaert et al.\ to these proceedings) is used here to derive the
ages, metallicities, luminosities and extinction for $\sim3\times10^4$
individual stars in the inner galactic Bulge, and the implications for the
structure and evolution of the galactic Bulge (cf.\ Wyse et al.\ 1997) are
discussed.

\section{Data and methods}

\begin{figure}
\plotfiddle{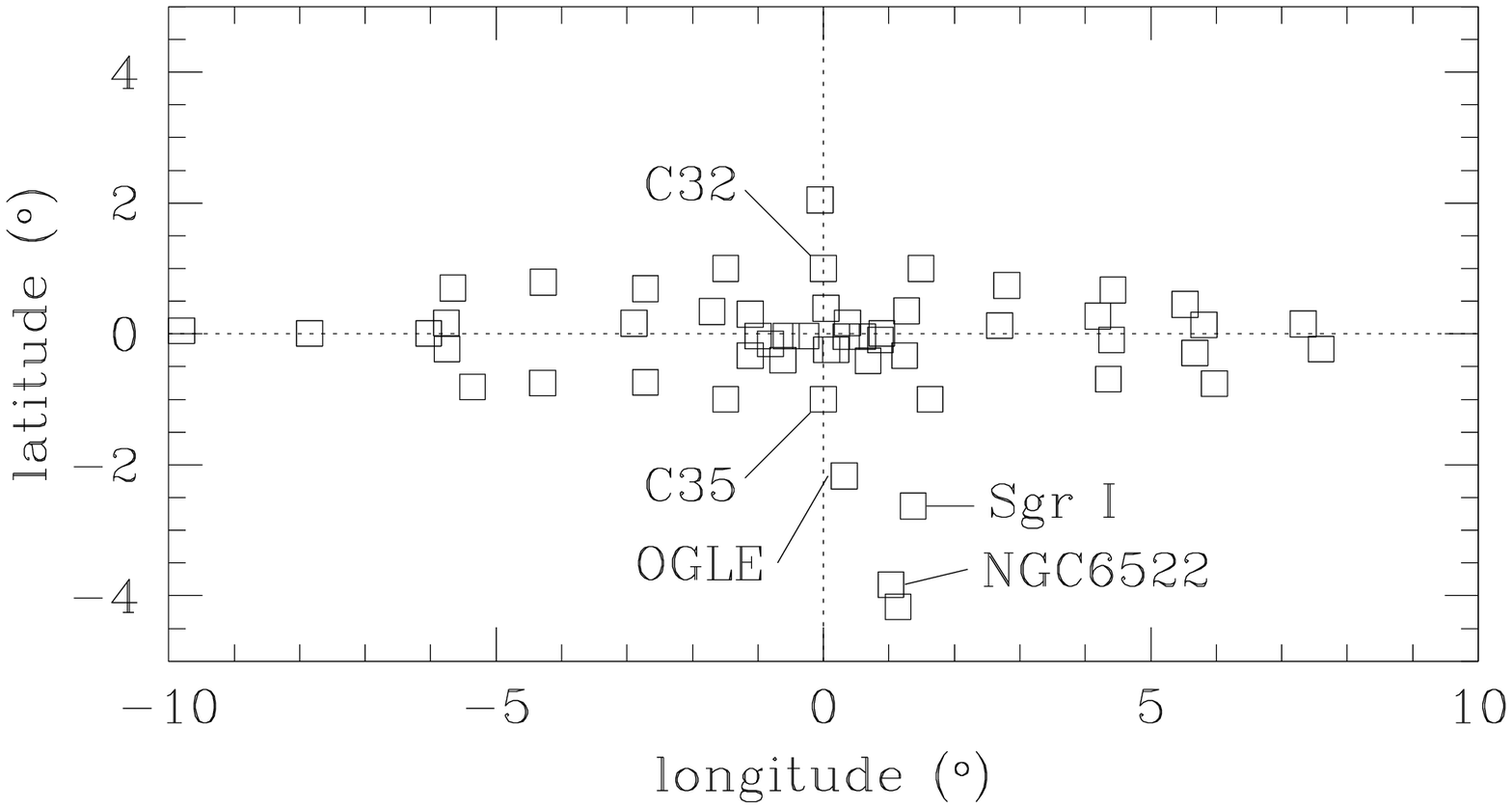}{58mm}{0}{60}{60}{-180}{-91}
\caption{Galactic map of the ISOGAL fields studied here.}
\end{figure}

The ISOGAL 7 and 15 $\mu$m images are typically $20^\prime\times20^\prime$, at
pixel scales of $3^{\prime\prime}$ or $6^{\prime\prime}$. Different filters
were used: LW2, LW5 and LW6 at 7 $\mu$m, and LW3 and LW9 at 15 $\mu$m. All
photometry has been transformed to the LW5 and LW9 filters (that best sample
the continuum) by applying small offsets. The fields cover regions along the
galactic Plane, as well as fields at higher latitude including the Large
Magellanic Cloud. The Milky Way is especially well-sampled for
$-10{\deg}<l_{\rm II}<+10{\deg}$ and $-2{\deg}{\la} b_{\rm II} {\la}+2{\deg}$,
which is the region studied here (Fig.\ 2). Preliminary mid-IR photometry for
point sources is complemented by DENIS I,J and K$_{\rm s}$-band photometry.

\begin{figure}
\plotfiddle{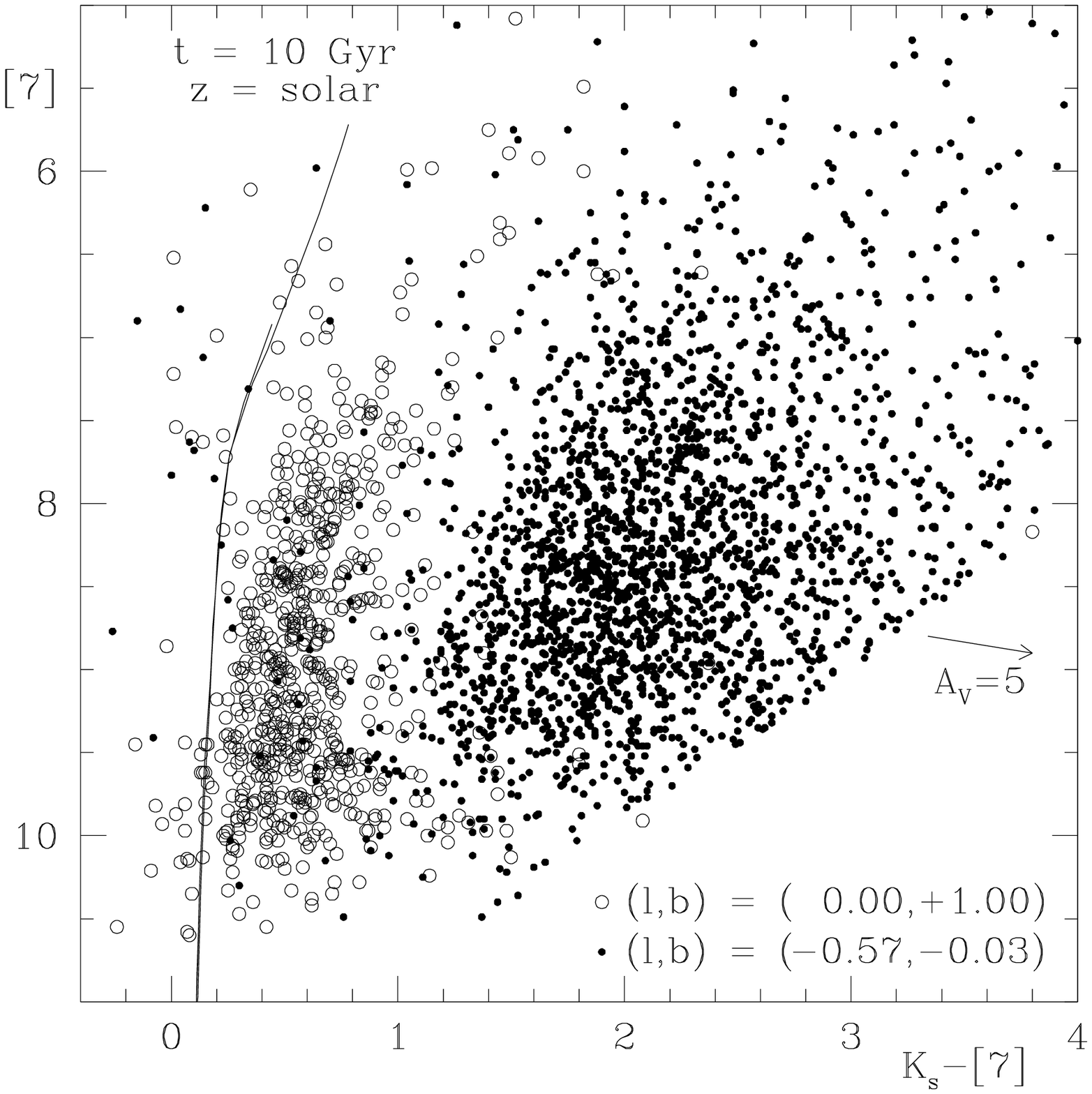}{111mm}{0}{60}{60}{-180}{-100}
\caption{The $[7]$ versus $(K-[7])$ diagram of the stellar populations in two
fields in the inner Galaxy that suffer from different amounts of (severe)
extinction. An isochrone for a 10 Gyr population of solar metallicity is
plotted for illustration.}
\end{figure}

An example of an IR colour-magnitude diagram for two fields is given in Fig.\
3, together with an isochrone for a 10 Gyr old population of solar metallicity
(Bertelli et al.\ 1994). The red colours of the stars are predominantly due to
severe interstellar extinction. The age, metallicity and extinction (adopting
Mathis 1990) are derived for each individual star by comparing its location
with respect to isochrones (Bertelli et al.\ 1994) in various IR
colour-magnitude diagrams, after proper computation of bolometric corrections
from a combination of Kurucz (1993) and MARCS~{\sc ii} (Fluks et al.\ 1994)
synthetic spectra. For stars without sufficient photometry to solve for all
three variables, the extinction is assigned as derived for neighbouring stars.
The luminosity and effective temperature are obtained as well. All stars are
assumed to be located at the distance of the galactic Centre, for which 8 kpc
is adopted.

\section{Interstellar extinction}

\begin{figure}
\plotfiddle{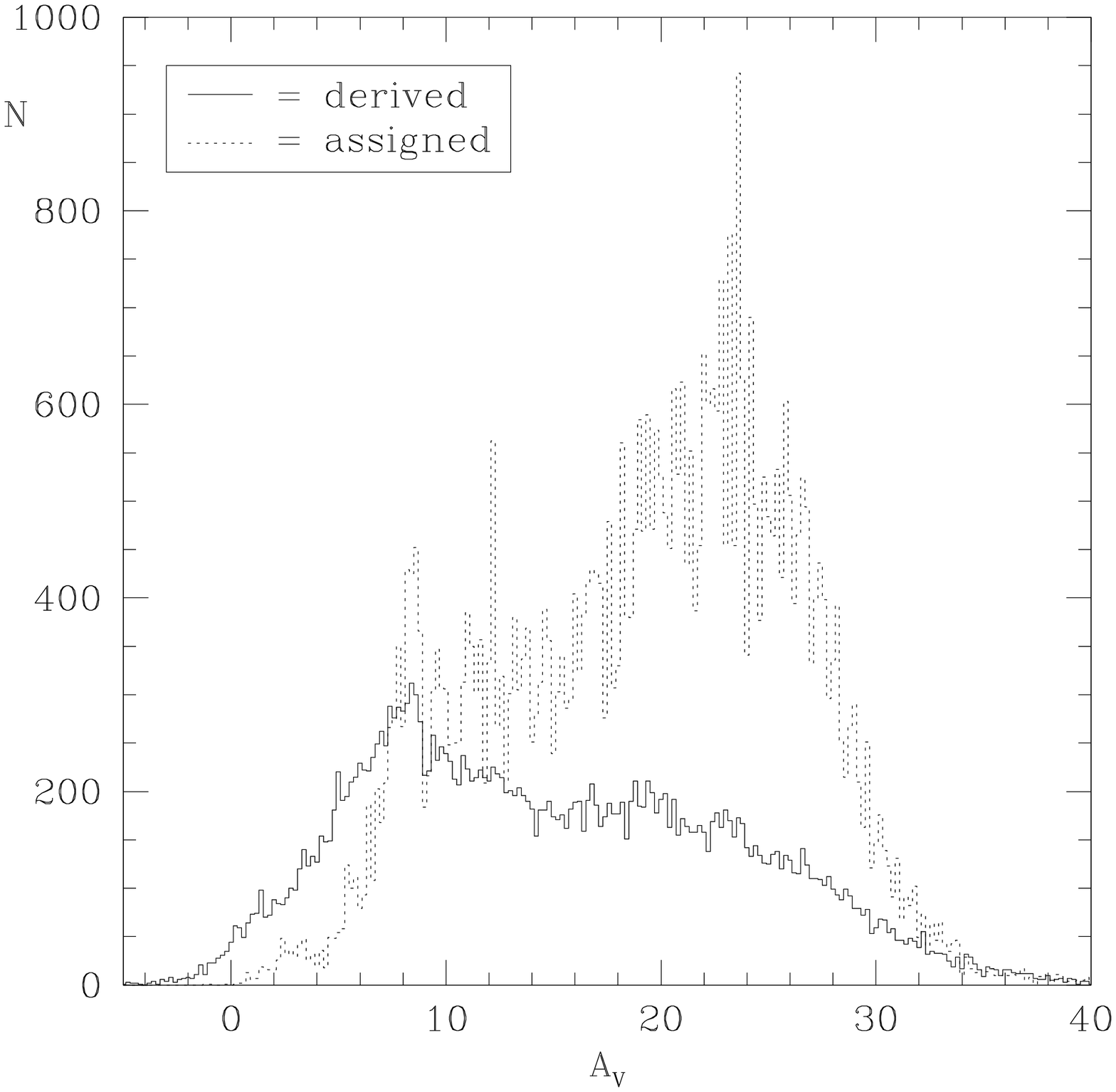}{108mm}{0}{60}{60}{-178mm}{-100mm}
\caption{Extinction distribution derived from the comparison of IR photometry
and isochrones. The dotted histogram shows the distribution of extinction
assigned to stars without sufficient photometry.}
\end{figure}

Derived extinction ranges from $A_{\rm V}\sim0$ in some areas at $b_{\rm
II}{\ga}3{\deg}$ to $A_{\rm V}>30$ mag in the galactic Plane at $|l_{\rm
II}|{\la}1{\deg}$ (Fig.\ 4). The extinction map of the inner galactic Plane
and Bulge based on DENIS photometry alone (Schultheis et al.\ 1999), although
unsurpassed in its spatial detail and depth, cannot quantify extinction of
$A_{\rm V}{\ga}20$ mag, which comprises half our sample of stars. The apparent
bimodality of the extinction distribution suggests at least two major
extinction components exist in the inner Galaxy. The stars without sufficient
photometry are predominantly found in regions of high extinction, as can be
seen in the distribution of the assigned extinction values.

\section{Stellar populations}

As a first result, the luminosity distribution of the stars is plotted in
Fig.\ 5. The Asymptotic Giant Branch (AGB) is detected even in the obscured
galactic core region. In most regions the Red Giant Branch (RGB) is detected
down to a few $10^2$ L$_\odot$. The derived effective temperatures are
generally $T_{\rm eff}\sim2500$ to 4000 K, confirming the red giant nature of
most of the stars. A few hundred bright mid-IR sources have been identified
that are interpreted as mass-losing AGB stars, some of which are associated
with OH maser emission (see also Omont et al.\ 1999; Glass et al.\ 1999; Ortiz
et al.\ 2000). The AGB stars thus make an excellent sample for the study of
the evolution and mass-loss of intermediate-mass populations, whilst the RGB
stars are ideally suited for the study of the early evolution and chemical
enrichment of the inner Galaxy.

\begin{figure}
\plotfiddle{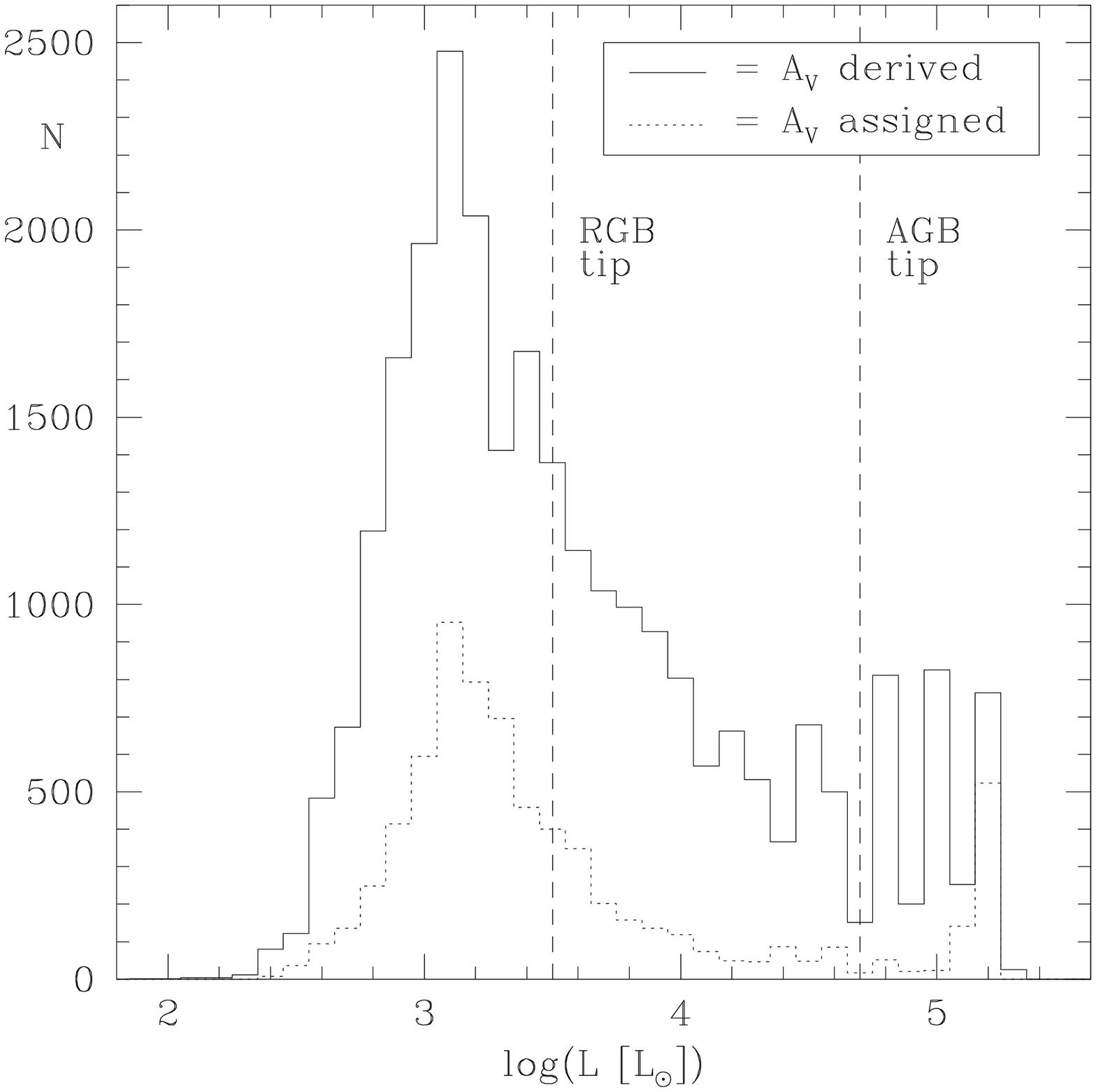}{108mm}{0}{60}{60}{-178mm}{-95mm}
\caption{Luminosity distribution derived from the comparison of IR photometry
and isochrones. All of the AGB, and a significant portion of the RGB are
detected.}
\end{figure}

\section{Ages and metallicities}

The metallicity distribution covers the wide range of $[M/H]{\in}[-1.7,0.4]$.
Although solar and super-solar metallicity stars amongst these are more
common, many of the stars have metallicities that are not much higher than of
stars in the galactic Halo. There is only a very marginal indication for a
negative metallicity gradient over the inner $\sim400$ pc from the galactic
Centre, suggesting that the inner galactic Bulge is rather --- but not
entirely --- homogeneous.

The dominant population of the inner Bulge is old, $t{\ga}10$ Gyr, but there
are indications for the presence of a significant intermediate-age population,
$t\sim1$ Gyr, and possibly an even younger population of $t{\la}100$ Myr too.
The age distribution is fairly uniform over the inner Bulge, although the
average age of the youngest component may slightly increase outwards of the
galactic Plane (traced up to $z=\pm400$ pc). Hence there is no strong
signature of an inner Disk component distinct from the inner Bulge.

This would suggest that in the galactic Bulge star formation did not switch
off completely after the first generations of stars were born and that we now
see as the old, metal-rich population. Instead, the intermediate-age
population of AGB stars would imply on-going star formation over most of the
galactic history, while the young population might represent a recent epoch of
enhanced star formation, possibly due to a minor merger event.

The derived ages should be taken with caution, though, until the final ISOGAL
photometry is analysed, and the results can be confirmed when spectroscopic
information is taken into account.

\begin{figure}
\plotfiddle{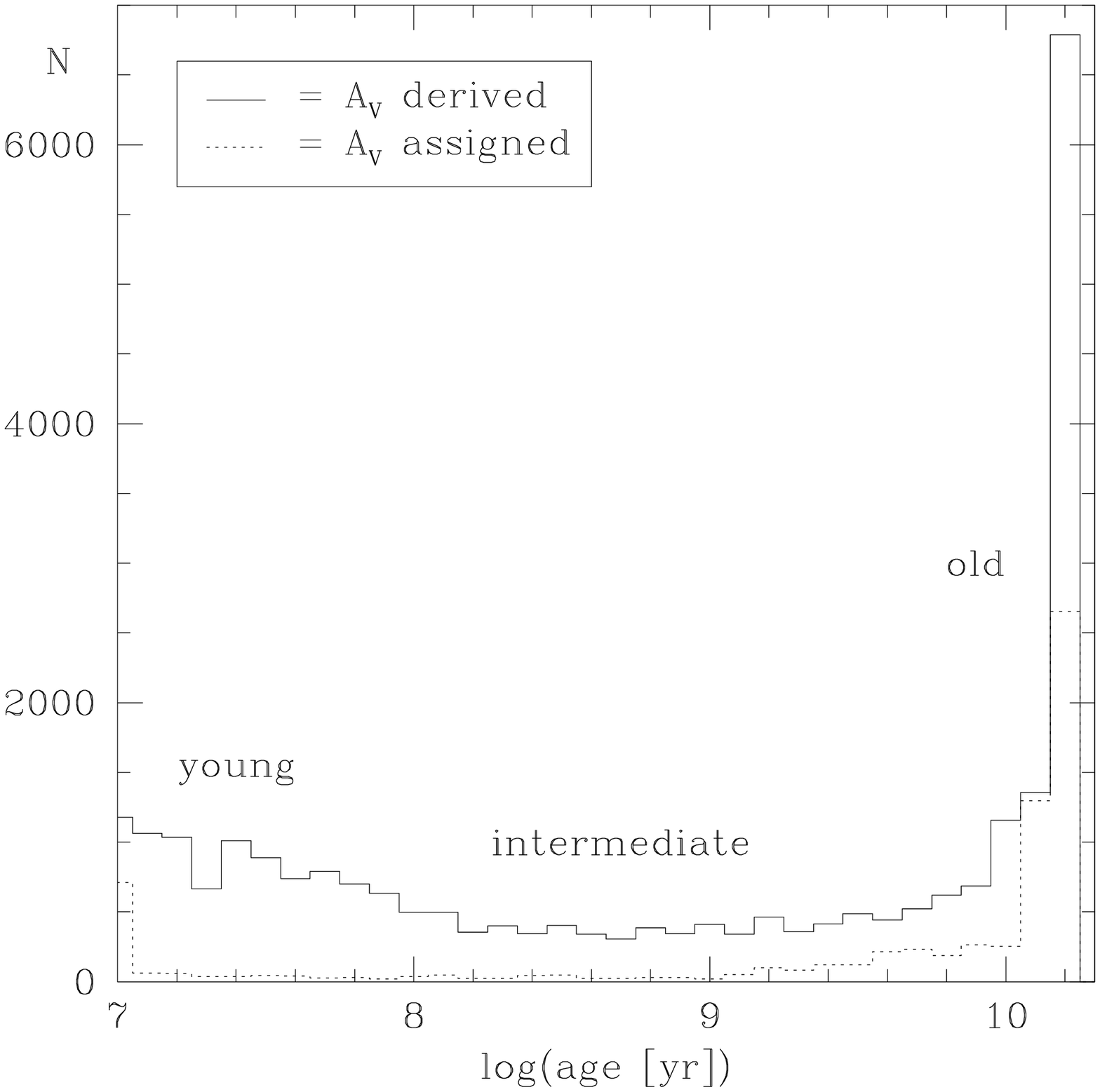}{108mm}{0}{60}{60}{-178mm}{-95mm}
\caption{Ages of inner Bulge stars derived from the comparison of IR
photometry and isochrones. Three populations are present: old ($t{\ga}10$
Gyr), intermediate ($t\sim1$ Gyr) and young ($t{\la}100$ Myr).}
\end{figure}

\section{Three-dimensional structure of the inner galactic Bulge}

The extinction-corrected luminosity function shows a clear asymmetry in the
galactic Plane at either side of the galactic Centre: it is brighter at
$l_{\rm II}\sim-6{\deg}$ than at $l_{\rm II}\sim+6{\deg}$ (Fig.\ 6). This can
be understood in terms of differential depth effects, if the inner Bulge has a
tri-axial shape on a radial scale of $R\sim1.4$ kpc under an angle
$i\sim53{\deg}$, with the near side at negative galactic longitude. This
geometry is much alike that proposed for the larger scale Bar, which is,
however, rather a disk phenomenon. Within the inner $|l_{\rm II}|<2{\deg}$
there is a hint of an asymmetry in the opposite sense. If confirmed, this
would make the Bulge resemble a spiral pattern instead. The structure of the
inner few kpc of the Galaxy thus seems to be highly complex. The analysis
presented here gives a preview of what future deep IR surveys may do in terms
of mapping the Milky Way in three dimensions. The results should be combined
with kinematic data to construct a complete and coherent picture of what the
inner Milky Way really looks like, before any firm conclusions can be drawn.

\begin{figure}
\plotfiddle{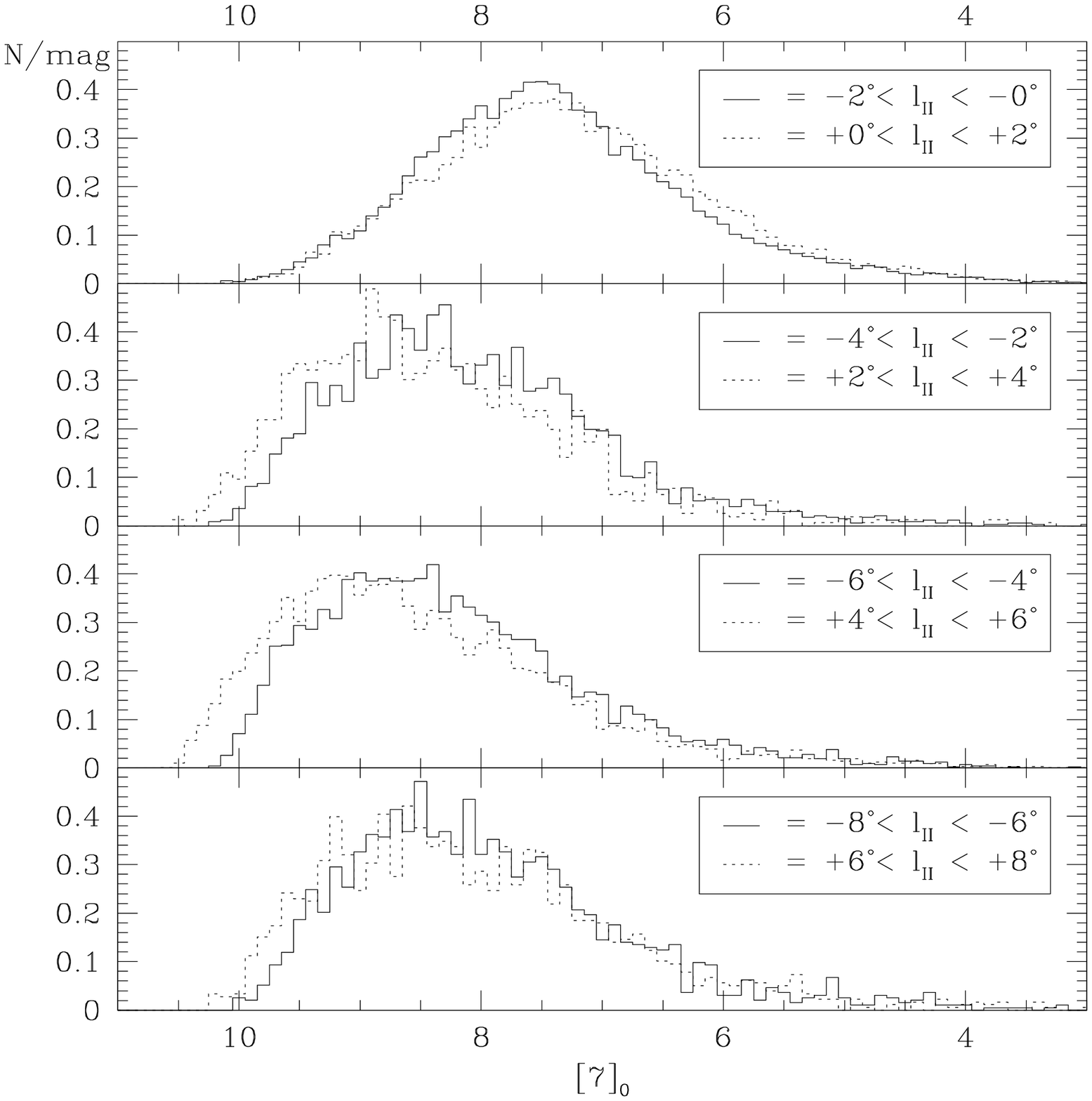}{108mm}{0}{60}{60}{-178mm}{-106mm}
\caption{Offsets between the extinction-corrected 7 $\mu$m luminosity
functions at either side of the galactic Centre can be traced across the inner
kpc of the galactic Plane, and may arise from a tri-axial Bulge.}
\end{figure}

\section{Future prospects}

The definitive results of the analysis presented here will make use of the
final ISOGAL photometric catalogue, which is envisaged to be completed in
Autumn 2000. Good quality spectroscopic data in the 0.6--2.3 $\mu$m range have
been obtained for $>200$ mass-losing AGB stars and other heavily obscured IR
objects in the innermost regions of the Galaxy. A deep near-IR survey with the
Cambridge IR Survey Instrument has just started at Las Campanas Observatory,
and will detect many more Bulge stars than with DENIS, including stars as
faint as the red clump. These new data will improve greatly our understanding
of the nature of the stellar populations, and better constrain their
metallicities and ages. The resulting view on the structure and evolution of
the inner parts of our own Milky Way galaxy may then serve as an important
template for the study of the structure and evolution of more distant
galaxies and the Universe as a whole.

\acknowledgments{I thank the organising committee for giving me the
opportunity to present this work at a rewarding conference in a wonderful
place. O anjinho Joana est\'{a} obrigado para tudo o seu apoio e
paci\^{e}ncia.}

\end{document}